\documentclass[fleqn]{article}
\usepackage{amsfonts}

\newcommand{\bls}[1]{\renewcommand{\baselinestretch}{#1}}

\def\noi{\noindent}

\makeatletter
\renewcommand{\section}{\@startsection{section}{1}{0pt}%
        {-3.5ex plus -1ex minus -.2ex}{2.3ex plus .2ex}%
        {\large\bf\protect\raggedright}}

\renewcommand{\subsection}{\@startsection{subsection}{2}{0pt}%
        {-3ex plus -1ex minus -.2ex}{1.4ex plus .2ex}%
        {\normalsize\bf\protect\raggedright}}

	\renewcommand{\@oddhead}{\raisebox{0pt}[\headheight][0pt]{%
   \vbox{\hbox to\textwidth{\rightmark \hfil \rm \thepage \strut}\hrule}}}
\renewcommand{\@evenhead}{\raisebox{0pt}[\headheight][0pt]{%
   \vbox{\hbox to\textwidth{\thepage \hfil \leftmark \strut}\hrule}}}
\newcommand{\heads}[2]{\markboth{\protect\small\it #1}{\protect\small\it #2}}

\makeatother

\def\prepno#1#2#3   {\thispagestyle{empty}
  \noi  \unitlength=1mm    \begin{picture}(178,10)
          \put(#3,15){\shortstack[r]{{\bf #1}\\ {\small\rm #2}}}
	 \end{picture}
 }
\newcommand{\Title}[1]{\noi {\uppercase{\Large #1}} \\}

\def\Aunames#1{\noi{\large\bf #1}}
\def\auth#1{${}^{#1}$}
\def\Addresses#1{\medskip\noi \protect
	\begin{description}\itemsep -3pt
        {\it #1} \end{description}}
\def\addr#1#2{\item[${}^{#1}$]{\it #2}}
\newcommand{\Abstract}[1]{\vskip 2mm \begin{center}
        \parbox{16.4cm}{\small\noi #1} \end{center}\medskip}

\newcommand{\email}[2]{\footnotetext[#1]{e-mail: #2}
		\addtocounter{footnote}{1}}

\newcommand{\Ref}[1]{Ref.\,\cite{#1}}

\def\nqq{\hspace*{-2em}}
\def\nhq{\hspace*{-0.5em}}

\def\cm{\hspace*{1cm}}

\def\ten#1{\mbox{$\cdot 10^{#1}$}}

\def\Jl#1#2{{\it #1\/} {\bf #2},\ }

\def\CQG#1 {\Jl{Clas. Qu. Grav.}{#1}}
\def\DAN#1 {\Jl{Dokl. AN SSSR}{#1}}
\def\GC#1 {\Jl{Grav. \& Cosmol.}{#1}}
\def\GRG#1 {\Jl{Gen. Rel. Grav.}{#1}}
\def\JETF#1 {\Jl{Zh. Eksp. Teor. Fiz.}{#1}}
\def\JETP#1 {\Jl{Sov. Phys. JETP}{#1}}
\def\JHEP#1 {\Jl{JHEP}{#1}}
\def\JMP#1 {\Jl{J. Math. Phys.}{#1}}
\def\NPB#1 {\Jl{Nucl. Phys.}{B\ #1}}
\def\NP#1 {\Jl{Nucl. Phys.}{#1}}
\def\PLA#1 {\Jl{Phys. Lett.}{#1A}}
\def\PLB#1 {\Jl{Phys. Lett.}{#1B}}
\def\PRD#1 {\Jl{Phys. Rev.}{D\ #1}}
\def\PRL#1 {\Jl{Phys. Rev. Lett.}{#1}}

\newcommand{\eqsection}{\makeatletter
	\@addtoreset{equation}{section}
	\renewcommand{\theequation}{\arabic{section}.\arabic{equation}}
	\makeatother}
\def\al{&\nhq}
\def\lal{&&\nqq {}}
\def\eq{Eq.\,}
\def\eqs{Eqs.\,}
\def\beq{\begin{equation}}
\def\eeq{\end{equation}}
\def\bear{\begin{eqnarray}}
\def\bearr{\begin{eqnarray} \lal}
\def\ear{\end{eqnarray}}
\def\earn{\nonumber \end{eqnarray}}

\def\nnn{\nonumber\\ \lal }

\def\yy{\\[5pt] {}}

\def\eql{\al =\al}

\def\tst{\textstyle}

\def\fract#1#2{{\tst\frac{#1}{#2}}}

\def\half{{\fract{1}{2}}}

\def\diag{\mathop{\rm diag}\nolimits}

\def\const{{\rm const}}

\def\DAL{\mathop{\raisebox{3.5pt}{\large\fbox{}}}\nolimits}

\def\GR{general relativity}

\def\mn{_{\mu\nu}}
\def\MN{^{\mu\nu}}
\def\mN{_\mu^\nu}

\def\cR{{\cal R}}

\def\rcr{\rho_{\rm cr}}

\bls{1.04}
\heads	{K.A. Bronnikov, V.N. Melnikov	and M. Novello}
{Possible time variations of $G$ in scalar-tensor theories of gravity}

\begin{document}
\twocolumn[
\prepno{gr-qc/0208028}{\GC{\rm Suppl. II} 18--21 (2002)}{115}

\Title{Possible time variations of $G$ \yy
	in scalar-tensor theories of gravity}

\Aunames
	{K.A. Bronnikov\auth{a,1}, V.N. Melnikov\auth{a,b,2}
			and Mario Novello\auth{b,3}}

\Addresses{
\addr a {VNIIMS, 3-1 M. Ulyanovoy Str., Moscow, 117313, Russia; \\
	Institute of Gravitation and Cosmology,
	Peoples' Friendship University of Russia,\\
	6 Miklukho-Maklaya St., Moscow 117198, Russia}
\addr b {Centro Brasileiro de Pesquisas Fisicas - CBPF/MCT,
	Rua Dr. Xavier Sigaud, 150,
	22290-180 - Rio de Janeiro, RJ, Brazil }
}

\Abstract{We estimate the possible variations of the gravitational constant
$G$ in the framework of a generalized (Bergmann-Wagoner-Nordtvedt)
scalar-tensor theory of gravity on the basis of the field equations, without
using their special solutions. Specific estimates are essentially related to
the values of other cosmological parameters (the Hubble and acceleration
parameters, the dark matter density etc.), but the values of $\dot G/G$
compatible with modern observations do not exceed $10^{-12}$. }

\medskip
] 
\email 1 {kb@rgs.mccme.ru}
\email 2 {melnikov@rgs.phys.msu.su, rgs@com2com.ru}
\email 3 {novello@cbpf.br}

\section{Introduction}

 Dirac's Large Numbers Hypothesis (LNH) is the origin of many theoretical
 explorations of time-varying $G$.  According to the LNH, the value of
 $\dot{G}/G$  should approximately coincide with the Hubble rate.
 Although it has become clear in the recent decades that the Hubble rate is
 too high to be compatible with experiment, the enduring legacy of Dirac's
 bold stroke is the acceptance by modern theories of non-zero values of
 $\dot{G}/G$ as being potentially consistent with physical reality.

 There are three problems related to $G$, whose origin lies mainly in
 unified model predictions:
1) absolute $G$ measurements,
2) possible time variations of $G$,
3) possible range variations of $G$, i.e., non-Newtonian, or new
interactions.  For 1) and 3) see \cite{Mel}.

After the original {\it Dirac hypothesis\/} some new concepts appeared
and also some generalized {\em theories\/} of gravitation admitting
variations of the effective gravitational coupling. We can single out three
stages in the development of this field:

\begin{enumerate} \itemsep -2pt
\item
Study of theories and hypotheses with variations of fundamental
physical constants, their predictions and confrontation with experiments
(1937-1977).

\item
Creation of theories admitting variations of an effective gravitational
constant in a particular system of units, analyses of experimental and
observational data within these theories \cite{Stan} (1977-present).

\item
Analyses of variations of fundamental physical constants within unified
models \cite{Mel} (present).
\end{enumerate}

Different theoretical schemes lead to temporal variations of the
effective gravitational constant:
\begin{enumerate}              \itemsep -2pt
\item
Empirical models and theories of Dirac type, where $G$ is simply replaced
with $G(t)$.

\item
Numerous scalar-tensor theories (STT) of Jordan-Brans-Dicke type, with
$G$ depending on the scalar field $\phi(t)$ or a number of scalar fields.

\item
Gravitational theories with a nonminimally coupled (in particular,
conformal) scalar field arising in different approaches \cite{Stan} (they
can actually be treated as special cases of STT).

\item
Multidimensional unified theories in which there are dilatonic
fields and effective scalar fields appearing in our 4-dimensional
spacetime from extra dimensions \cite{Mel}. They may also help one
in solving the problem of a variable cosmological constant
from Planckian to present values and the cosmic coincidence problem.
\end{enumerate}

 A striking feature of the present status of theoretical physics is that
 there is no satisfactory theory unifying all four known interactions;
 most modern unification theories do not admit unique and universal constant
 values of physical constants and of the Newtonian gravitational coupling
 constant $G$ in particular. In this paper we discuss the bounds that
 may be suggested by a general class of STT. One can mention that STT are
 among the viable alternatives to \GR; on the one hand, they are widely used
 for comparison with observations and, on the other, their different
 versions emerge in the field limits of the candidate ``theories of
 everything''.

 Although the bounds on $\dot G$ and $G(r)$ are in some classes of theories
 rather wide on purely theoretical grounds, since any theoretical model
 contains a number of adjustable parameters, we note that observational data
 concerning other phenomena, in particular, cosmological data, place limits
 on the possible ranges of these adjustable parameters.

 Here we restrict ourselves to the problem of $\dot G$ (for $G(r)$
 see [1--4]). We show that various theories predict the value of
 $\dot{G}/G$ to be $10^{-12}/$yr  or less. The significance of this fact for
 experimental and observational determinations of the value of or upper
 bound on $\dot G$ is the following: any determination with error bounds
 significantly better than $10^{-12}/$yr (combined with experimental bounds
 on other parameters) will typically be compatible with only a small portion
 of existing theoretical models and will therefore cast serious doubt on the
 viability of all other models. In short, a tight bound on $\dot G$, in
 conjunction with other astrophysical observations, will be a very effective
 ``theory killer'' and/or significantly reduce the class of viable theories.
 Any step forward in this direction will be of utmost significance.

 Some estimations of $\dot G$ had been done long ago in the framework of
 general scalar-tensor and multidimensional theories using the values of
 cosmological parameters ($\Omega$, $H$, $q$ etc) known at that time
 \cite{Stan, Mel, we-NC, Z}. It is easy to show that for modern values they
 predict $\dot{G}/G$ at the level of $10^{-12}/$yr and less (see also recent
 estimations of A. Miyazaki \cite{Mi}, predicting time variations of $G$ at
 the level of $10^{-13} {\rm yr}^{-1}$ for a Machian-type cosmological
 solution in the Brans-Dicke theory).

 The most reliable, by now, experimental bounds on $\dot{G}/G$ (spacecraft
 radar ranging \cite{Hel}) and lunar laser ranging \cite{Dic}) give a
 limit of $10^{-12}/$yr), so any results at this level or better will be
 very important for solving the fundamental problem of variations of
 constants and for discriminating between viable unified theories. So,
 realization of such multipurpose new generation type space experiments like
 Satellite Energy Exchange (SEE) for measuring $\dot G$ and also absolute
 value of $G$ and Yukawa type forces at the ranges of metres and the Earth
 radius \cite {san,ref2} become extremely topical.

\section{Scalar-tensor cosmology and variations of $G$}

   We are going to estimate the order of magnitude of variations of the
   gravitational constant $G$ due to cosmological expansion in the framework
   of scalar-tensor theories (STT) of gravity, using modern data on the
   cosmological parameters.

   Consider the general (Bermann-Wagoner-Nordtvedt) class of STT where
   gravity is characterized by the metric $g\mn$ and the scalar field
   $\phi$; the action is
\bearr
    S = \int d^4 x \sqrt{g}\bigl\{ f(\phi) \cR[g]            \label{act}
\nnn \cm
        + h(\phi)g\MN\phi_{,\mu}\phi_{,\nu} -2 U(\phi) + L_m \bigr\}.
\ear
   Here $\cR[g]$ is the scalar curvature,
   $g = |\det (g_{\mu \nu})|$; $f,\ h$ and $U$ are certain functions of
   $\phi$, varying from theory to theory, $L_m$ is the matter Lagrangian.

   This formulation of the theory corresponds to the Jordan conformal frame,
   in which matter particles move along geodesics and hence the weak
   equivalence principle is valid, and non-gravitational fundamental
   constants do not change. In other words, this is the frame well
   describing the existing laboratory, geophysical and cosmological
   observations.

   Among the three functions of $\phi$ entering into (\ref{act})
   only two are independent since there is a freedom of transformations
   $\phi = \phi(\phi_{\rm new})$. We use this arbitrariness, choosing
   $h(\phi) \equiv 1$, as is done, e.g., in \Ref{star00}.
   Another standard parametrization is to put $f(\phi)=\phi$ and
   $h(\phi) = \omega(\phi)/\phi$ (the Brans-Dicke parametrization of the
   general theory (\ref{act})). In our parametrization $h\equiv 1$, the
   Brans-Dicke function $\omega(\phi)$ is  $\omega(\phi) = f/f_\phi^2$; here
   and henceforth, the subscript $\phi$ denotes a derivative with respect to
   $\phi$. The Brans-Dicke STT is the particular case $\omega = \const$, so
   that in (\ref{act})
\beq
    f(\phi) =    \phi^2/(4\omega), \cm h \equiv 1.     \label{BD}
\eeq
   The field equations that follow from (\ref{act}) read
\bearr
   \DAL \phi - \half \, \cR \, f_\phi + U_\phi = 0,       \label{Ephi}
\\ \lal
    f(\phi) \Bigl(\cR\mN - \half \delta\mN \cR\Bigr)
   = -\phi_{,\mu} \phi^{,\nu}
    + \half\delta\mN \phi^{,\alpha}\phi_{,\alpha}      \label{EE}
\nnn \cm
      - \delta\mN U(\phi) + (\nabla_{\mu}\nabla^\nu - \delta\mN \DAL)f
      - T\mN{}_{\rm (m)},
\ear
   where $\DAL$ is the D'Alembert operator, and the last term in (\ref{EE})
   is the energy-momentum tensor of matter.

   Consider now isotropic cosmological models with the standard FRW metric
\bearr
   ds^2 = dt^2 -a^2(t)\biggl[\frac{dr^2}{1-kr^2} + r^2         \label{ds}
           (d\theta^2 + \sin^2\theta\,d\varphi^2)\biggr],
\nnn
\ear
   where $a(t)$ is the scale factor of the Universe, and $k=1,\ 0,\ -1$
   for closed, spatially flat and hyperbolic models, respectively.
   Accordingly, we assume $\phi = \phi(t)$ and the energy-momentum tensor of
   matter in the perfect fluid form
   $T\mN{}_{\rm (m)} = \diag(\rho, -p, -p, -p)$
   ($\rho$ is the density and $p$ is the presuure).

  The field equations in this case can be written as follows
  (the dot denotes $d/dt$:
\bearr
    \ddot\phi + 3\frac{\dot a}{a} \dot{\phi}                    \label{Ef}
    -\frac{3}{a^2} (a\ddot{a} + \dot{a}{}^2 +k) + U_{\phi} =0,
\\ \lal                                                          \label{E0}
     \frac{3f}{a^2}(\dot{a}{}^2+k)
            = \half \dot\phi{}^2 +U -3\frac{\dot a}{a} \dot f +\rho,
\\ \lal
    \frac{f}{a^2}(2a\ddot a + \dot{a}{}^2 +k)                    \label{E1}
        = -\half \dot\phi{}^2 +U-\ddot f-2\frac{\dot a}{a} \dot f -p.
\ear

  To connect these equations with observations, let us fix the cosmic time
  $t$ at the present epoch (i.e., consider the instantaneous values of all
  quantities) and introduce the standard observables:

  $H=\dot a/a$ (the Hubble parameter),

  $q=-a\ddot a/\dot a{}^2$ (the deceleration parameter),

  $\Omega_m = \rho/\rcr$ (the matter density parameter),

\noi
   where $\rho_{\rm cr}$ is the critical density, or,
   in our model, the r.h.s. of \eq (\ref{E0}) in case $k=0$:
   $\rcr = 3fH^2$. This is slightly different from the usual
   definition $\rcr = 3H^2/8\pi G$ where $G$ is the Newtonian
   gravitational constant. The point is that the locally measured Newtonian
   constant in STT differs from $1/(8\pi f)$; provided the derivatives
   $U_{\phi\phi}$ and $f_{\phi\phi}$ are sufficiently small, one has
   \cite{star00}
\beq                                                          \label{Geff}
    8\pi G_{\rm eff} = \frac{1}{f} \frac{2\omega +4}{2\omega+3}.
\eeq
   (more details can be found in Refs.\,\cite{MR1,MR2} where the connection
   between $G_{\rm eff}$ and $\omega$ was studied on the basis of
   cosmological solutions with local inhomogeneities and the equations of
   particle motion.)

   Since, according to the solar-system experiments, $\omega\geq 2500$,
   for our order-of-magnitude reasoning we can safely put $8\pi G = 1/f$,
   and, in particular, our definition of $\rcr$ now coincides with the
   standard one.

   The time variation of $G$, to a good approxiamtion, is
\beq
    \dot G/G \approx - \dot f/f = gH,                     \label{Gdot}
\eeq
   where, for convenience, we have introduced the coefficient $g$ expressing
   $\dot G/G$ in terms of the Hubble parameter $H$.

  \eqs (\ref{Ef})--(\ref{E1}) contain too many arbitrary parameters for
   making a good estimate of $g$. Let us now introduce some restrictions
   according to the current state of observational cosmology:

\medskip\noi
   {\bf (i)} $k=0$ (a spatially flat cosmological model, so that the total
       density of matter equals $\rcr$);

\medskip\noi
   {\bf (ii)} $p=0$ (the pressure of ordinary matter is negligible compared
       to the energy density);

\medskip\noi
   {\bf (iii)} $\rho = 0.3\,\rcr$ (the ordinary matter, including its dark
       component, contributes to only 0.3 of the critical density;
       unusual matter, which is here represented by the scalar field,
       comprises the remaining 70 per cent).

  Then \eqs (\ref{E0}) and (\ref{E1}) can be rewritten in the form
\bear
    \half \dot\phi{}^2 + U - 3H\dot f \eql 2.1 H^2 f,    \label{E0'}
\\
    -\half \dot\phi{}^2 + U - 2H\dot f -\ddot f \eql (1-2q) H^2 f.
                                    \label{E1'}
\ear
   Subtracting (\ref{E1'}) from (\ref{E0'}), we exclude the ``cosmological
   constant'' $U$, which can be quite large but whose precise value is
   hard to estimate. We obtain
\beq
     \dot\phi{}^2 - H\dot f + \ddot f = (1.1 + 2q) H^2 f.    \label{*}
\eeq

  The first term in \eq (\ref{*}) can be represented in the form
\[
     \dot\phi{}^2 = \dot f{}^2 (df/d\phi)^{-2} = \dot f{}^2 \omega/f,
\]
   and $\dot f/f$ can be replaced with $-gH$. The term $\ddot f$ can be
   neglected for our estimation purposes. To see this, let us use as an
   example the Brans-Dicke theory, in which $f= \phi^2/(4\omega)$.
   We then have
\[
   \ddot f = (\dot\phi{}^2 + \phi\ddot\phi)/(2\omega);
\]
   here the first term is the same as the first term in \eq (\ref{*}), times
   the small parameter $1/(2\omega)$. Assuming that $\phi\ddot\phi$ is of
   the same order of magnitude as $\dot\phi{}^2$ (or only slightly greater),
   we see that, generically, $|\ddot f| \ll \dot\phi{}^2$. Note that our
   consideration is not restricted to the Brans-Dicke theory and concerns
   the model (\ref{act}) with an arbitrary function $f(\phi)$ and
   an arbitrary potential $U(\phi)$.

   Neglecting $\ddot f$, we see that (\ref{*}), divided by $H^2 f$, leads
   to a quadratic equation with respect to $g$:
\beq
    \omega g^2 +g - q' =0,                                \label{**}
\eeq
   where $q' = 1.1 +2q$.

   According to modern observations, the Universe is expanding with an
   acceleration, so that the parameter $q$ is, roughly, $-0.5\pm 0.2$,
   hence we can take $|q'| \leq 0.4$. (Note that this condition is only
   plausible rather than certain.)

   In case $q'=0$ we simply obtain $g = -1/\omega$. Assuming
\[
   H = h_{100}\cdot 100\ {\rm km}/({\rm s.Mpc})
	    \approx h_{100} \cdot 10^{-10}\ {\rm yr}^{-1}
\]
   and $\omega\geq 2500$, we come to the estimate
\beq
     |\dot G/G| \leq 4\ten{-14} h_{100}\ {\rm yr}^{-1},       \label{est1}
\eeq
   where $h_{100}$ is, by modern views, close to 0.7. So (\ref{est1})
   becomes
\beq
     |\dot G/G| \leq 3\ten{-14}\ {\rm yr}^{-1}.            \label{est1'}
\eeq

   For nonzero values of $q'$, solving the quadratic equation (\ref{**}) and
   assuming $q'\omega\gg 1$, we arrive at the estimate
   $|g| \sim \sqrt{q'/\omega}$, so that, taking $q'=0.4$ and again
   $\omega\geq 2500$, we have instead of (\ref{est1})
\beq
     |\dot G/G| \leq 1.3\ten{-12} h_{100}\ {\rm yr}^{-1}
		\approx 0.9\ten{-12} {\rm yr}^{-1},           \label{est2}
\eeq
   where we have again put $h_{100}=0.7$.

   We conclude that, in the framework of the general STT, modern
   cosmological observations, taking into account the solar-system data,
   restrict the possible variation of $G$ to values within $10^{-12}$/yr.
   This estimate may be considerably tightened if the matter density
   parameter $\Omega_m$ and the (negative) deceleration parameter $q$ will
   be determined more precisely.

   Our estimates are rather universal since they do not use special
   solutions to the field equations, but actually rest on the well-justified
   assumption that the expansion of the Universe occurs without abrupt
   changes in its parameters during a fairly long period before now.

\section{Discussion}

   Summarizing the above considerations, we can conclude that
   restrictions on possible nonzero values of $\dot{G}$ give no
   bound on the possible class of generalized gravitation theories, but
   in the framework of some fixed theory any restriction on $\dot{G}$
   restricts the possible class of models.

   We note that similar estimations of $\dot G$ can be made for different
   multicomponent multidimensional models \cite{J}, giving a result on the
   level of $10^{-12}$/yr and less for, e.g., dust and p-brane matter
   sources.

   We can also mention that the behaviour of the gravitational constant can
   actually be much more complex and intriguing than a simple time (or even
   range) dependence: very recently, there appeared two papers, which may
   open a new series of theoretical and experimental studies related to
   possible anisotropy in the absolute value of $G$ \cite{K}
   and/or its possible dependence on the latitude and longitude of the
   laboratory where $G$ was measured \cite{ML}.

\end{document}